\documentclass[prb,twocolumn,groupedaddress,showpacs]{revtex4}

\usepackage{graphicx}
\usepackage{graphics}
\graphicspath{{/home/tfs/and/Thoughts/Words/Brus/}{/home/tfs/and/Thoughts/Words/Brus/Img/}}
\newcommand{\ket}[1]{{\left | {#1} \right\rangle}}
\newcommand{\bra}[1]{{\left\langle {#1} \right |}}

\begin{document}
\title{Full frequency voltage noise spectral density of a single electron transistor.}
\author{Andreas K\"ack}
\affiliation{Microtechnology Center at Chalmers MC2, Department of Microelectronics
and Nanoscience, Chalmers University of Tecnology and G\"oteborg University,
S-412 96, G\"oteborg, Sweden}
\author{G\"oran Johansson}
\affiliation{Institut f\"ur Theoretische Festk\"orperphysik, Universit\"at Karlsruhe, D-761 28 Karlsruhe, Germany}

\author{G\"oran Wendin}
\affiliation{Microtechnology Center at Chalmers MC2, Department of
  Microelectronics and Nanoscience, Chalmers University of Tecnology
  and G\"oteborg University, S-412 96, G\"oteborg, Sweden}

\begin{abstract}
We calculate the full frequency spectral density of voltage fluctuations in a
Single Electron Transistor (SET), used as an electrometer biased above the
Coulomb threshold so that the current through the SET is carried by sequential
tunnel events. We consider both a normal state SET and a superconducting SET.
The whole spectrum from low frequency telegraph noise to quantum noise at
frequencies comparable to the SET charging energy $(E_{C}/\hbar)$,
and high frequency Nyquist noise is described. We take the energy exchange
between the SET and the measured system into account using a real-time diagrammatic
Keldysh technique.
The voltage fluctuations determine the back-action of the SET onto the measured
system and we specifically discuss the case of superconducting charge qubit read-out
and measuring the so-called Coulomb staircase of a single Cooper pair box.
\end{abstract}
\maketitle

\section{Introduction}

Solid-state realizations of qubits are of real interest due to the possibility of using
lithographic techniques to integrate the large number of qubits, needed for a fully
functional QC. 
The Single Electron Transistor (SET) has been suggested
as a read-out device for different solid-state charge qubits 
\cite{Aassime,Kane,e_on_He,Averin,MakhlinPRL}.

Aassime et al.~\cite{Aassime} have shown
that the Radio-Frequency SET~\cite{RFSET} (RF-SET)
may be used for single shot read-out
of the Single Cooper-pair Box (SCB) qubit~\cite{SCB_Bouchiat,Nakamura}.
This is possible if the measurement time $t_{ms}$ needed to resolve
the two states of the qubit is much shorter than
the time $t_{mix}$ on which the qubit approaches its new steady-state
determined by the back-action due to voltage fluctuations on the SET.
In a previous short 
paper\cite{JohanssonPRL} we provided further support for 
this result by calculating the full frequency voltage noise spectral density
of the SET, including the effect of energy exchange between the qubit
and the SET.

In this paper we 
give a full account of 
 the calculation as well as a
thorough discussion of the effect of back-action in measuring the
so-called Coulomb staircase of an SCB qubit. We also include a section
about the back-action from a superconducting SET.

Since the qubit is carefully shielded from all unwanted interactions
with its environment it is reasonable to assume that the back-action
from the SET charge measurement is the dominating noise source, even
though the two systems are only weakly coupled. This further motivates
choosing measurements on a charge qubit to discuss the spectral properties
of SET back-action, compare e.g. Ref.\cite{Aguado}. In non-qubit systems, e.g. a normal state single-electron
box, other sources of dissipation dominate over the back-action from the
charge measurement. Furthermore we do not discuss the dephasing of the
qubit induced by the presence of the SET\cite{DevoretNature,MakhlinRMP}.
Although it is a very important subject, the dephasing time is mainly
determined by the zero frequency fluctuations, and in this limit our
result coincides with previous expressions\cite{Korotkov_spectral}.

The structure of the paper is the following: In section \ref{sec:SCPB} we
discuss the basic properties of the SCB qubit and the effect of gate
voltage fluctuations. Furthermore the difference between asymmetric and symmetric definitions of
noise spectral density is noted and also the connection to mixing-time and the Coulomb staircase.
In section \ref{sec:SETmodel} the model for the SET is introduced and in section \ref{sec:Diagrams}
the real-time diagrammatic Keldysh technique is described and the expression for the
spectral density of voltage fluctuations on the SET island is derived.
Section \ref{sec:back-action} describes the properties of the voltage fluctuations
in different frequency regimes, the effect on the mixing-time and the Coulomb staircase
both for a normal state SET and a superconducting SET.

\section{The single Cooper-pair box qubit}
\label{sec:SCPB}
The qubit is here made up of the two lowest lying energy levels in a
single Cooper-pair box (SCB) \cite{SCB_Bouchiat}. 
An SCB is a small superconducting island
coupled to a superconducting reservoir via a Josephson junction.
The Hamiltonian of the system can be written in the charge basis as\cite{MakhlinRMP}
\begin{equation}
  \label{eq:qubitH}
  H_{q}=\sum_n 4E_{qb} (n-n_g)^2\ket{n}\bra{n}-\frac{E_J}{2}\left[\ket{n}\bra{n+1}+\ket{n+1}\bra{n}\right],
\end{equation}
where $E_{qb}=e^2/C_{qb}$ is the charging energy, $n$ is the number of extra Cooper-pairs on the island, $E_J$ is
the Josephson energy of the junction and
$n_g=C^{qb}_{g} V^{qb}_g/2e$ is the number of gate-induced Cooper-pairs. In order
to get a good Cooper-pair box we need $k_B T\ll E_J \ll E_c <
\Delta$, where $\Delta$ is the superconducting gap and $T$ is the
temperature. The low temperature is required to prevent thermal
excitations and the high superconducting gap is needed to suppress 
quasiparticle tunneling.
The eigenenergies now form parabolas, varying with the gate voltage.
\begin{figure}
  \begin{center}
    \includegraphics[width=9cm]{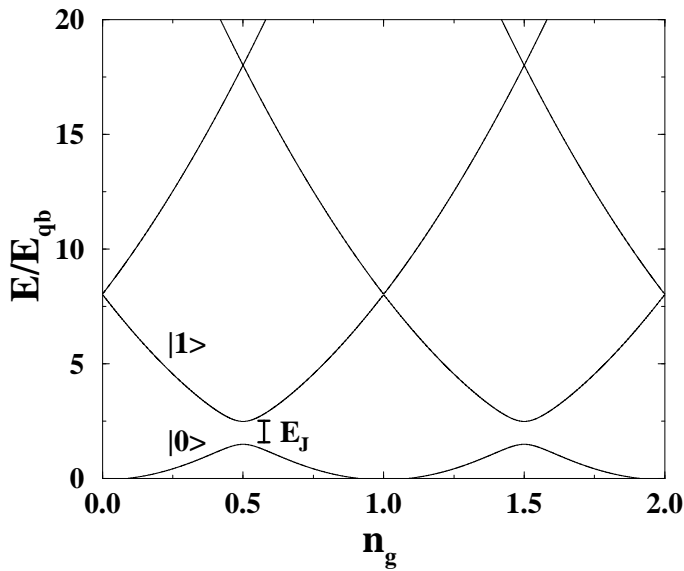}
    \caption{The energy bands of an SCB. The states $\ket{0}$ and
      $\ket{1}$ denote the eigenstates on the island. The quasiparticle branches have been left out for
      simplicity. The gap is due to hybridization of the charge states
      by the Josephson coupling.}
    \label{fig:qbands}
  \end{center}
\end{figure}
For suitable values of the gate voltage (close to $n_g=1/2$), 
and for $E_J \ll E_c$, the system 
reduces to an effective two level system as the two lowest lying
charge states are well separated from the states with higher energy.
Including only the two lowest
lying states in Eq.~(\ref{eq:qubitH}) the qubit Hamiltonian becomes
\begin{equation}
  \label{eq:qHreduced}
  H_{q}=-\frac{4E_{qb}}{2} (1-2n_g)\sigma_z -\frac{E_J}{2}\sigma_x,
\end{equation}
where $\sigma_{x,z}$ are the Pauli matrices (and the states
$\ket{\uparrow} = \left(\begin{array}{c} 1\\0\end{array}\right)$ and
$\ket{\downarrow} = \left(\begin{array}{c} 0\\1\end{array}\right)$
correspond to zero and one extra Cooper-pair on the qubit island).
By changing the gate voltage the eigenstates of the qubit can be tuned from being
almost pure charge states to a superposition of charge
states. 

The ground state/first excited state of the system, written in the
charge basis ($E_J=0$) are
\begin{eqnarray}
  \label{eq:eigenstates}
  \ket{0}&=&\cos(\eta/2)\ket{\uparrow}+\sin(\eta/2)\ket{\downarrow}\nonumber,\\
  \ket{1}&=&-\sin(\eta/2)\ket{\uparrow}+\cos(\eta/2)\ket{\downarrow},
\end{eqnarray}
where $\eta=\arctan(E_J/4E_{qb}(1-2n_g))$ is the mixing angle. 
The energy difference between the two states is \mbox{$\Delta
  E=\sqrt{(4E_{qb})^{2}(1-2n_g)^2+E_{J}^{2}}$} and the average
charge of the eigenstates is
\begin{eqnarray}
  \label{eq:charge}
  Q_{0}=  2 e
  |\left< \downarrow|0\right>|^2=2e\sin^2(\eta/2)\nonumber,\\
  Q_{1}=  2 e| \left<\downarrow|1\right>|^2=2e\cos^2(\eta/2).\nonumber\\
\end{eqnarray}

\subsection{Qubit transitions induced by SET voltage fluctuations}
When the SET is turned on in order to measure the qubit, the
voltage fluctuations on the SET-island induce 
a fluctuating charge on the qubit island. This is equivalent
to a fluctuating qubit gate charge $n_g \rightarrow n_g+\delta n_g(t)$,
giving rise to a fluctuating term in the qubit Hamiltonian, written in the charge basis as
\begin{equation}
  \label{eq:voltagefluct}
  \delta H_{q}(t)=\frac{4 E_{qb}}{2}2 \delta n_g(t)\sigma_z=2e\kappa\delta V(t)\sigma_z,
\end{equation}
where $\kappa=C_c/C_{qb}$.
Here $\delta V(t)$ represents the voltage fluctuations on the SET-island, and 
we have neglected a term
quadratic in $\delta n_g(t)$. In the qubit eigenbasis, using the rotation defined by
Eq.~(\ref{eq:eigenstates}), the fluctuations in Eq.~(\ref{eq:voltagefluct}) become
\begin{equation}
  \label{eq:10}
  \delta H(t)=2e \kappa \delta V(t) \left[\cos(\eta)\sigma_z+\sin(\eta)\sigma_x\right].
\end{equation}

The fluctuating voltage on the SET island can induce
transitions between the eigenstates of the qubit.
If the capacitive coupling to the SET is small ($\kappa \ll 1$) we can use the Fermi golden rule 
to calculate the transition rates:
\begin{eqnarray}
  \label{eq:7}
  \Gamma_{rel}(\Delta E)&=&\frac{e^2}{\hbar^2}\frac{E_{J}^2}{\Delta E^2}\kappa^2
  S_V(\Delta E/\hbar), \\
  \Gamma_{exc}(\Delta E)&=&\frac{e^2}{\hbar^2}\frac{E_{J}^2}{\Delta E^2}\kappa^2  S_V(-\Delta E/\hbar),
\end{eqnarray}
where $\Gamma_{rel}$ is the relaxation rate and $\Gamma_{exc}$ is
excitation rate and $S_V(\Delta E)$ is the asymmetric (see Eq.~(\ref{eq:asymmSv})) spectral density of the
voltage fluctuations on the SET island.
The fraction \mbox{$E_J/\Delta E=\sin(\eta)$} comes
from Eq.~(\ref{eq:10}) as it is only $\sigma_x$ that causes any 
transitions between the states. Note that the
$\cos(\eta)\sigma_z$ term causes fluctuation of the energy levels, 
leading to phase fluctuations and dephasing.

\subsection{Asymmetric noise - Coulomb staircase}
In our calculations we emphasize the energy exchange between the qubit
and the SET and separate between the contributions from processes leading to
the qubit loosing energy and the contribution from processes leading
to the qubit gaining energy (or equivalently: the SET absorbing or emitting energy).
Because of this, we will maintain this separation of the noise
spectral density of the SET into contributions from positive and
negative frequencies and therefore use the asymmetric expression for
the voltage fluctuations
\begin{equation}
\label{eq:asymmSv}
S_V(\omega)=\int_{-\infty}^{\infty}e^{-i \omega \tau}
\langle\delta V(\tau)\delta V(0)\rangle .
\end{equation}

As we are primarily interested in the processes in the SET we chose our
reference so that positive frequencies correspond to the SET absorbing
energy and negative frequencies correspond to the SET emitting energy.

One example where the separation is necessary is in describing the
back-action of the SET while measuring the so-called Coulomb
staircase, i.e. the average charge of the qubit as a function of its gate voltage.
In an ideal situation with no energy available from an
external source, at zero temperature, the qubit would follow the ground state adiabatically
and the charge would increment in steps of $2e$ at $n_g=n+0.5$, $n$ integer. These steps are not
perfectly sharp because of the Josephson energy mixing the charge
states. This mixing of charge states results in a maximal derivative given by $4E_C/E_J$\cite{SCB_Bouchiat}.
When adding a noise source, e.g. by increasing the temperature or attaching a
noisy measurement device, the steps will be rounded further due to a finite
population of the excited state.

Assuming that the SET is the dominant noise source, and that the two state
approximation of the qubit is valid, the steady-state population of the qubit
is given by

\begin{equation}
  \label{eq:steadyqubitup}
    P_{1}^{st,qb} = \Gamma_{exc}(\Delta E)/[\Gamma_{exc}(\Delta
    E)+\Gamma_{rel}(\Delta E)]
\end{equation}
\begin{equation}
     \label{eq:steadyqubitdown}
    P_{0}^{st,qb} = \Gamma_{rel}(\Delta
    E)/[\Gamma_{exc}(\Delta E)+\Gamma_{rel}(\Delta E)].
\end{equation}
The corresponding expression for the average charge is then given by
\begin{equation}
  \label{eq:coulomb}
  Q(\Delta E) = Q_{0}(\Delta E) P_{0}^{st,qb}(\Delta
  E)+Q_{1}(\Delta E) P_{1}^{st,qb}(\Delta E),
\end{equation}
where $Q_{\uparrow/\downarrow}$ is the charge of the excited
state/ground state defined in Eq.~(\ref{eq:charge}). If we set $P_{1}^{st,qb}=0$ and
$P_{0}^{st,qb}=1$, we recover the ideal Coulomb staircase, as
this corresponds to the system following the ground state
adiabatically.

\subsection{Symmetric noise - Mixing time}
For quantities that depend on the summed rate of relaxation- 
and excitation-processes in the SET the separation of
absorption and emission might not be necessary, and the symmetrised
expression for voltage fluctuations
\begin{equation}
S^{sym}_V(\omega)=\int_{-\infty}^{\infty} d\tau e^{-i \omega \tau}
\langle\delta V(\tau)\delta V(0)+\delta V(0)\delta V(\tau)\rangle
\end{equation}
can be used. Note that $S^{sym}_V(\omega)=S_V(\omega)+S_V(-\omega)$.

One example is the time it takes the qubit to reach its steady-state,
after the SET is switched on. This time is called the mixing time,
and the information about the initial state-population is lost on this
timescale. For weak coupling it is \cite{MakhlinRMP,DevoretNature}

\begin{equation}
  \label{eq:mixtime}
  \frac{1}{\tau_{mix}}=\Gamma_{rel}(\Delta E)+\Gamma_{exc}(\Delta E)
\propto S^{sym}_V(\Delta E/\hbar).
\end{equation}

\section{SET model}

\label{sec:SETmodel}
\begin{figure}
  \begin{center}
    \includegraphics[width=8cm]{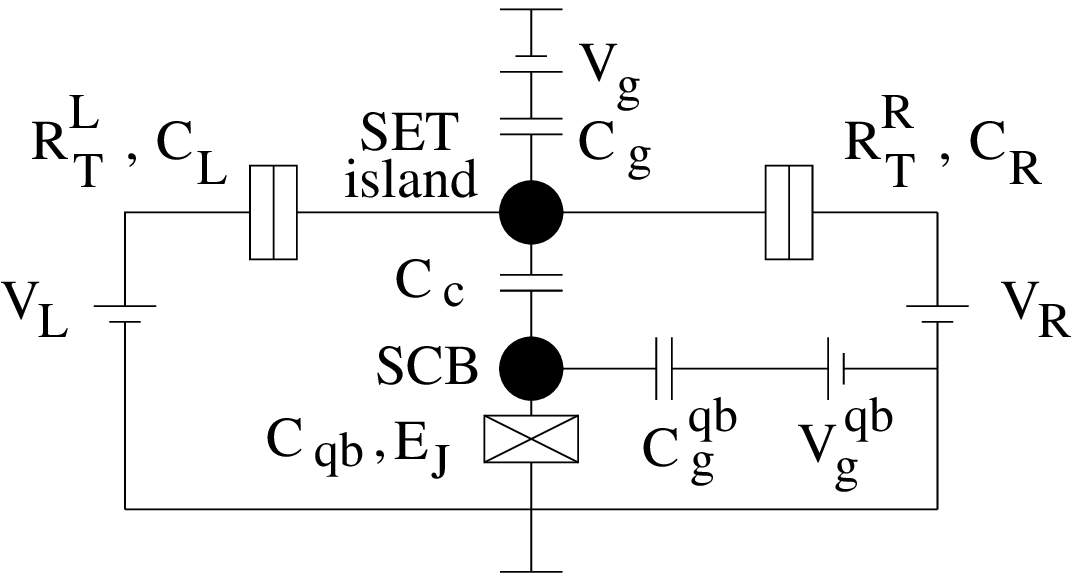}
    \caption{The SET.}    
  \end{center}
\end{figure}

We follow the outline of Ref.~\cite{SchoellerPRB} and model
the SET by the Hamiltonian
\begin{equation}
  H=H_L+H_R+H_I+V+H_T=H_0+H_T,
\end{equation}
where
\begin{equation}
  H_r=\sum_{kn}\epsilon^r_{kn}a^\dagger_{krn}a_{krn},
  \qquad
  H_I=\sum_{ln}\epsilon_{ln} c^\dagger_{ln} c_{ln}
\end{equation}
describe noninteracting electrons in the 
left/right lead ($H_r, r\in\{L,R\}$) and on the island ($H_I$).
The quantum numbers $n$ denote transverse channels including spin,
and $k, l$ denote momenta. The Coulomb interaction on the
island is described by
\begin{equation}
  V(\hat{N})=E_C(\hat{N}-n_x)^2,
\end{equation}
where $\hat{N}$ denotes the excess number operator,
\mbox{$E_C=e^2/2C$} the charging energy (\mbox{$C=C_L+C_R+C_g+C_c$}),
$n_x$ the fractional number of electrons induced by the
external voltages
($n_x$ is the fractional part of $(C_L V_L + C_R V_R + C_g V_g)/e$)
and $e$ the electron charge. The tunneling term is
\begin{eqnarray}
  H_T&=&\sum_{r=L,R}\sum_{kln}(T^{rn}_{kl}\,a^\dagger_{krn}c_{ln}
  e^{-i\hat{\Phi}}\,+\,T^{rn*}_{kl}\,c^\dagger_{ln}a_{krn}
  e^{i\hat{\Phi}}) \nonumber\\ &=& H^{-}+H^{+},
\end{eqnarray}
                                
where the operator $e^{\pm i\hat{\Phi}}$ changes the excess
particle number on the island by $\pm 1$ and $T^{rn}_{kl}$
are the tunneling matrix elements. $\hat{\Phi}$ is the canonical
conjugate to $\hat{N}, ([\hat{\Phi},\hat{N}]=i)$. In this case of
a metallic island containing a large number of electrons, the charge
degree of freedom $N=0,\pm1,...$ is to a very good approximation
independent of the electron degrees of freedom $l, n$. The  
terms $ H^{+}$ and $H^{-}$ represent electron tunneling to and from 
the SET island. The form of the tunneling terms (with the 
$e^{i\hat{\Phi}}$ term) is a consequence of separating state space 
into electron $l,n$ and charge $N$ degrees of freedom. This is 
also reflected in the partitioning of the density matrix introduced 
below.

The spectral density of voltage fluctuations on the SET island is
described by the Fourier transform of the voltage-voltage correlation
function
\begin{eqnarray}
  \label{noisedef}
  S_V(\omega) &=& \frac{e^2}{C^2} \int_{-\infty}^\infty d\tau
  e^{-i\omega\tau} \mathrm{Tr}\left\{\rho_{st}(t_0) \delta \hat{N}(\tau) \delta
    \hat{N}(0)\right\}=\nonumber\\
  &=&\frac{e^2}{C^2}\int_{-\infty}^\infty d\tau
  e^{-i\omega\tau} \mathrm{Tr}\left\{\rho_{st}(t_0) \left(\hat{N}(\tau) \hat{N}(0) -
      \bar{N}^2\right)\right\},\nonumber\\
\end{eqnarray}
where $\bar{N}$ is the average of $\hat{N}$. The $\bar{N}^2$-term in
Eq.~(\ref{noisedef}) assures that the correlation function vanishes for
large $\tau$. In Fourier space this term does not contribute at finite
frequency, and at zero frequency it compensates for the steady-state
delta function. In order to simplify our expressions we leave this
term out, and keep the frequency finite during the calculations.

In Eq.~(\ref{noisedef}) $\rho_{st}(t_0)$ is the density matrix of the system
in the steady-state, which is assumed to have been reached
at some time $t_0$ before the fluctuation occurs ($t_0 < \min\{0,\tau\}$).
$\rho_{st} = \rho^e_{eq} \otimes \rho^c_{st} $ is the tensor product of the equilibrium (Fermi distributed)
density matrix $\rho^e_{eq}$ for the electron degrees of freedom in each
reservoir ($L, R, I$) and a reduced density matrix $\rho^c_{st}$,
describing the charge degrees of freedom. Since the tunneling events 
between the SET island and the electrodes are incoherent due to the
low conductance of the tunnel junctions, the charge
states will be incoherent, and $\rho^c_{st}$ is  therefore taken to be
diagonal~\cite{SchoellerNATO} with elements $P^{st}_N$ denoting the
steady-state  probability of being in charge state $N$.

\section{Diagrammatic treatment}
\label{sec:Diagrams}
\begin{figure}
  \begin{center}
    \includegraphics[width=7cm]{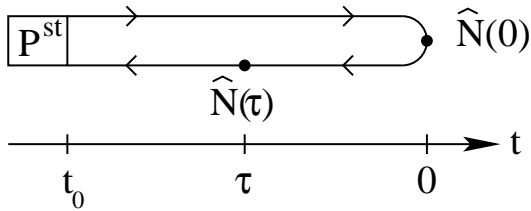}
    \caption{The propagation of the unperturbed steady state density matrix. This
    corresponds to the first term in Eq.~(\ref{eq:explicitperturbation}).\label{fig:timeloop}}
\end{center}
\end{figure}

We now expand the correlation function in Eq.~(\ref{noisedef}) in a
perturbation series in terms of the tunneling Hamiltonian ($H_T$)
along a Keldysh time contour (see Fig.~\ref{fig:timeloop}).
The trace over the electron degrees of freedom is then evaluated using Wick's theorem, which is possible
since the tunneling Hamiltonian is only bilinear in the fermionic
operators. 
Rewriting Eq.~(\ref{noisedef}) in the interaction picture gives (excluding 
the $\bar{N}^2$-term)
\begin{widetext}
\begin{equation}
  S_{V}(\omega)=\frac{e^2}{C^2}\int_{-\infty}^{\infty} d\tau e^{-i\omega\tau}
  \mathrm{Tr}\left\{\rho(t_0)S(t_0,\tau) \hat{N}(\tau) S(\tau,0) \hat{N}(0) S(0,t_0)\right\},
  \label{eq:noise_interaction}
\end{equation}
where $S(t_2,t_1)$ is the $S$-matrix that brings the system from the
time $t_1$ to time $t_2$, i.e. for $t_2 >t_1$
\begin{equation}
  \label{eq:s-matrix}
  S(t_2,t_1)=e^{-iT\int_{t_1}^{t_2} dt H_T(t)} = 1-i\int_{t_1}^{t_2}
  d\tau_1
  H_T(\tau_1)+(-i)^2\int_{t_1}^{t_2}d\tau_1\int_{t_1}^{\tau_1}d\tau_1 d\tau_2  H_T(\tau_1)H_T(\tau_2)+\dots,
\end{equation}
and analogously for $t_2 < t_1$. Defining $T$ as the time-ordering
operator along the Keldysh contour we can write
Eq.~(\ref{eq:noise_interaction}) as
\begin{eqnarray}
  S_{V}(\omega) &=& \frac{e^2}{C^2}\int_{-\infty}^{\infty} d\tau
  e^{-i\omega\tau}
  \left[\mathrm{Tr}\left\{\rho(t_0)\hat{N}(\tau)\hat{N}(0)\right\}-i\int_{-\infty}^{0}d\tau_1\mathrm{Tr}\left\{T\rho(t_0)\hat{N}(\tau)\hat{N}(0)H_T(\tau_1)\right\}
  \right.\nonumber\\
  && \left.+\frac{(-i)^2}{2!} \int_{-\infty}^{0}d\tau_1 \int_{-\infty}^{0}d\tau_2
    \mathrm{Tr}\left\{T\rho(t_0)\hat{N}(\tau)\hat{N}(0)H_T(\tau_1)H_T(\tau_2)\right\} +\dots \right] .  
  \label{eq:explicitperturbation}
\end{eqnarray}
\end{widetext}

Since the unperturbed Hamiltonian does not contain any couplings between
the leads and the island nor between the leads themselves, their degrees of
freedom are independent.
Moreover, as the trace of independent degrees of freedom is equal to the 
product of the respective traces and every perturbation term contains one
operator from one of the leads and one operator from the island, only
terms containing an even number of perturbation terms will contribute.

Using the diagrammatic language of Ref.~\cite{SchoellerPRB}, 
the noise correlation function $S_{V}(\omega)$ can be given a 
diagrammatic formulation.

\begin{figure}
  \begin{center}
    \includegraphics[width=7cm]{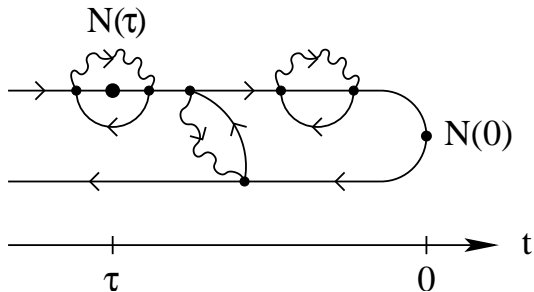}
    \caption{An example of a specific diagram. The wiggly lines
      correspond to island lines and the solid lines correspond to
      lead lines. The rightmost part correspond to an electron-hole
      excitation while the leftmost part corresponds to a correction to
      the external vertex. The middle part corresponds to a tunneling
      event as only the diagrams connecting the upper and lower branch
      changes the number of extra charges on the island.\label{fig:completeexample}}
  \end{center}
\end{figure}

Fig.~\ref{fig:timeloop} shows a diagrammatic representation 
of the first term in Eq.~(\ref{eq:explicitperturbation}) with zero $H_T$
perturbations, involving the 
evolution of the state described by the density matrix itself along the 
Keldysh time contour.
This process is represented by a solid line starting and ending at the
the density matrix, including the charge (fluctuation) operators
$\hat{N}(\tau)$ and $\hat{N}(0)$ marked by dots.
This gives the expectation value (statistical average)
of the charge fluctuation correlation (in the "ground state") in the absence
of any transport process.

\begin{figure}
  \begin{center}
    \includegraphics[width=6cm]{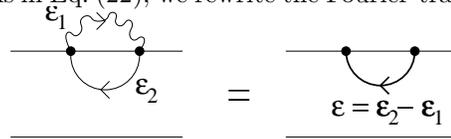}
    \caption{Two internal electron lines with reversed start and end
      points. They can be replaced by a single line with an energy
      equal to the difference in energy between the two corresponding
      to an electron-hole excitation.}    
    \label{fig:firstexample}
  \end{center}
\end{figure}

The first contributions to charge transport come from the third term in
Eq.~(\ref{eq:explicitperturbation}) with one $H^{+}$ and one $H^{-}$ perturbation, describing tunneling onto the SET island and back, changing
the charge state $N$ to $N\pm 1$. A diagrammatic representation of
this would be the middle part of Fig.~\ref{fig:completeexample}. Since the charge-transfer process can be viewed as 
an electron-hole excitation, creating a hole on an electrode and an 
electron on the island,  
in Fig.~\ref{fig:completeexample} there are new additional lines with arrows representing
electron-hole propagators (excitations).
Every internal time will form a vertex and the
propagator $\langle Ta(\tau_1)a^{\dagger}(\tau_2) \rangle$ will form a line going
from $\tau_2$ to $\tau_1$.
In this case with macroscopic metallic reservoirs with many transverse channels, 
the main contributing terms\cite{SchoellerPRB} will appear in combinations of
\begin{eqnarray}
  \label{eq:two2one}
  \langle T\,H^{+}(\tau_1)H^{-}(\tau_2)
  \rangle&=&\sum_{r_1}\sum_{r_2}\sum_{k_{1}l_1n_1}\sum_{k_2l_2n_2}
   \left[T_{k_1 l_1}^{r_1 n_1 *}T_{k_2 l_2}^{r_2 n_2} \nonumber\right.\times\\&&\times\langle T  
  a_{k_1r_1n_1}(\tau_1)a_{k_2r_2n_2}^{\dagger}(\tau_2) \rangle\times
  \nonumber\\&&\left.\times  \langle T c_{k_1r_1n_1}^{\dagger}(\tau_1)c_{k_2r_2n_2}(\tau_2)\rangle\right]\nonumber,\\
\end{eqnarray}
which  means that there will
always be pairs of internal lines with reversed start and end
points, one being a reservoir propagator and one being an island 
propagator, as shown in Fig.~\ref{fig:firstexample}. These
line-pairs are replaced by a single line, corresponding to an
electron-hole excitation, with an energy equal to the
difference in energy between the two lines.

In order to facilitate the evaluation of the time-ordered diagrams in
Eq.~(\ref{eq:explicitperturbation}), we rewrite
the Fourier-transform in Eq.~(\ref{noisedef}) as

\begin{widetext}
\begin{eqnarray}
  S_V(\omega) &=& \frac{e^2}{C^2} \int_{-\infty}^\infty d\tau
  e^{-i\omega\tau} \mathrm{Tr}\{\rho_{st}(t_0) \hat{N}(\tau) \hat{N}(0)\}
  = \frac{e^2}{C^2}\int_{-\infty}^{0} d\tau  e^{-i\omega\tau} \mathrm{Tr}\{\rho_{st}(t_0)
  \hat{N}(\tau) \hat{N}(0)\}\nonumber\\
  &&+\frac{e^2}{C^2}\int_{-\infty}^{0} d\tau  e^{+i\omega\tau} \mathrm{Tr}\{\rho_{st}(t_0)
  \hat{N}(0) \hat{N}(\tau)\} 
  = 2 \frac{e^2}{C^2} \mathrm{Re}\left[ \int_{-\infty}^{0} d\tau  e^{-i\omega\tau} \mathrm{Tr}\{\rho_{st}(t_0)
  \hat{N}(\tau) \hat{N}(0)\} \right],
\label{eq:newS}
\end{eqnarray}
\end{widetext}
where we have used that the steady-state is time invariant.
Furthermore we fix the 
specific time ordering of all internal and external times in all
diagrams. This make the diagrams straightforward to evaluate in the
frequency domain as all integrals thus become recursive Laplace-transforms.

Returning to Eq.~(\ref{eq:explicitperturbation}) and drawing all the
diagrams of the lowest non-trivial order we can divide them into two
categories: Dressings of the propagator $\Pi_{N,N^\prime}(\omega)$ that takes the system from the
charge state $N$ to the state $N^\prime$ (see
Fig.~\ref{fig:irreducible}), and vertex correction (see
Fig.~\ref{fig:all_vertex}).

\begin{widetext}
\begin{center}
\begin{figure}
  \begin{center}
    \includegraphics[width=10cm]{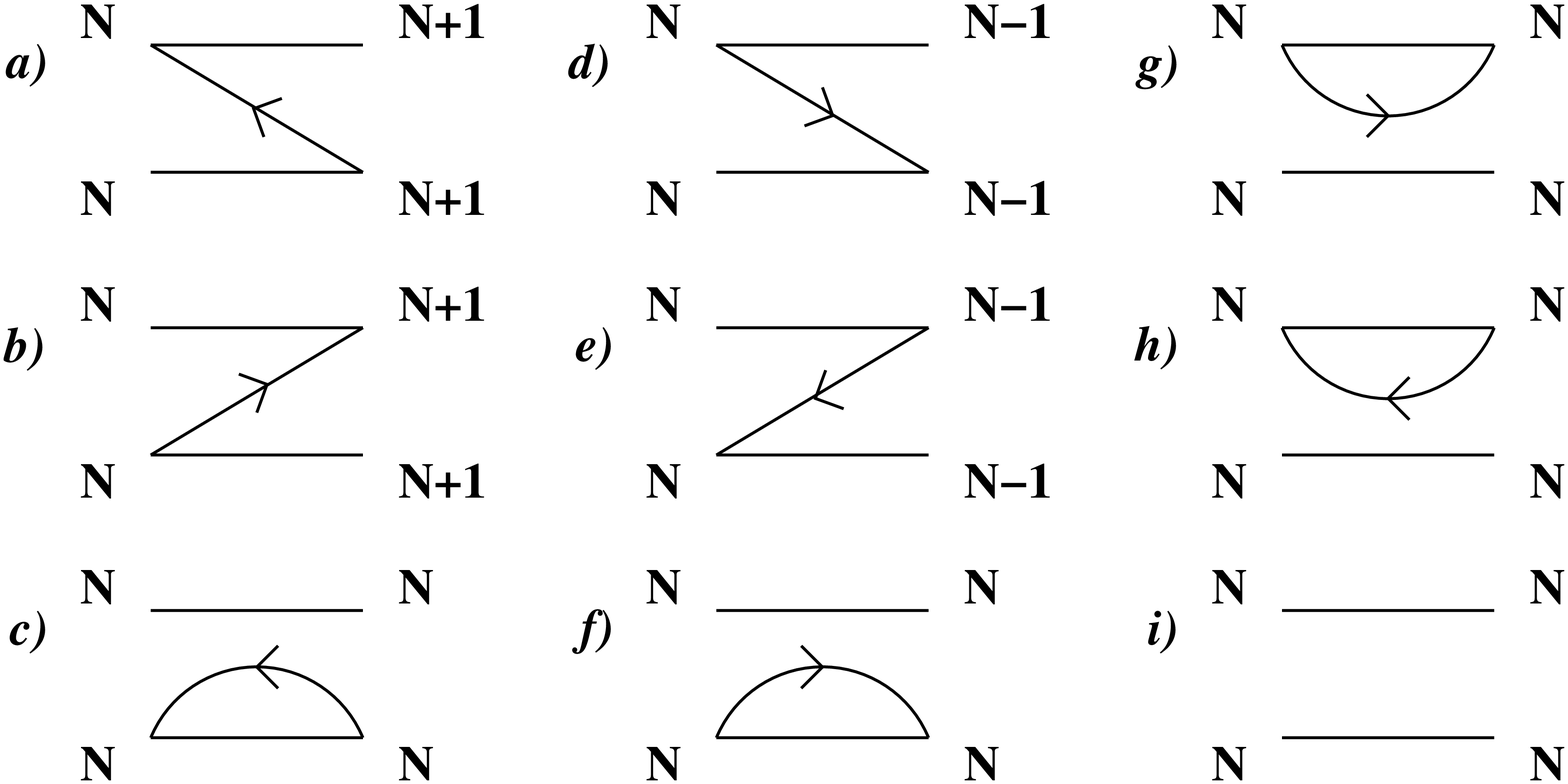}
    \caption{All the diagrams that enter the propagator, to the lowest
      order.}
    \label{fig:irreducible}
  \end{center}
\end{figure}
\end{center}
\end{widetext}

\begin{figure}
  \begin{center}
    \includegraphics[width=7cm]{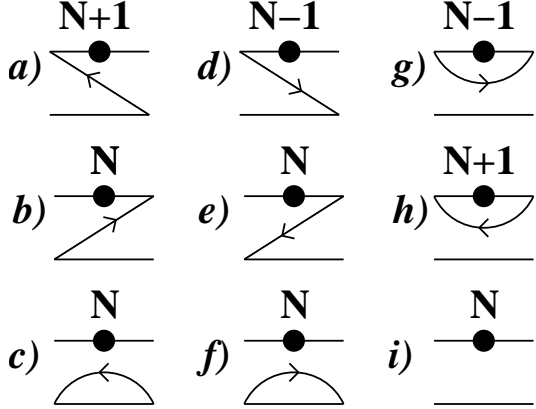}
    \caption{All the vertex corrections of the external vertex
      $\hat{N}(\tau)$, denoted by a dot,
      within the sequential tunneling approximation. Note the
      correspondence to Fig.~\ref{fig:irreducible}.}
    \label{fig:all_vertex}
  \end{center}
\end{figure}

Calculating for instance the diagram in Fig.~\ref{fig:irreducible}) using the
diagrammatic technique outlined in Ref.\cite{SchoellerPRB} yields 
\begin{equation}
  D_N = -\sum_{r}\lim_{\eta\rightarrow 0^{+}}\frac{i}{\pi}\int dE
  \frac{\gamma^{+}_{r}(E)}{\hbar\omega+E-\Delta_{N}+i\eta},
  \label{eq:irreduciblediagram}
\end{equation}
where $\Delta_{N}=V(N+1)-V(N)$ is the charging energy cost of
moving one electron from one of the leads onto the island when
there are $N$ extra electrons there. The factor $\gamma^{+}_{r}(\varepsilon)$ is
the inelastic (Golden Rule) tunneling rate through junction $r$ for electrons tunneling to the
island (correspondingly $\gamma_{r}^{-}(\varepsilon)$ is the inelastic tunneling rate of
electrons tunneling from the island through junction $r$)
given by the expression 
\begin{widetext}
\begin{equation}
  \gamma^{\pm}_{r}(\varepsilon)=\frac{\pi}{\hbar}\sum_{n} \int dE \rho(\varepsilon+E-
  eV_{r})\rho(E)
  f^{\pm}(\varepsilon+E-eV_r)f^{\mp}(E) |T^{rn}|^2
  \label{eq:gammarp},
\end{equation}
\end{widetext}
where $\rho$ is the density of states in the
reservoir $r$ and on the island, both assumed to be either
superconductors or normal metal, $f^{+}(E)$ is the Fermi-distribution,
\mbox{$f^{-}(E)=1-f^{+}(E)$}, $T^{rn}$ is the tunneling matrix element,
here assumed to be independent of energy and $V_r=(\mu_r-\mu_I)/e$ is the
voltage bias across the $r$ junction. 
Separating Eq.~(\ref{eq:irreduciblediagram}) into real and imaginary parts we get
\begin{widetext}
\begin{equation}
  D_N = -\sum_{r}\lim_{\eta\rightarrow 0^{+}}\frac{i}{\pi}\int dE
  \gamma^{+}_{r}(E)\left[\frac{\hbar\omega+E-\Delta_{N}}{(\hbar\omega+E-\Delta_{N})^2+\eta^2}-\frac{i\eta}{(\hbar\omega+E-\Delta_{N})^2+\eta^2}\right].
  \label{eq:real_im_diagram}
\end{equation}
\end{widetext}
The real part of the integral is small in the tunnel limit where the conductance 
is small, so we neglect these renormalization
effects (see Ref.~\cite{SchoellerPRB}) and concentrate on the imaginary part,
which gives
\begin{equation}
D_N = -\sum_{r}\gamma^{+}_{r}(\Delta_N-\hbar\omega).
\end{equation}
Introducing the notation
\begin{eqnarray}
  \label{eq:newrates}
  \gamma^{+}_N (\omega)=\sum_{r=R,L}\gamma^{+}_{r}(\Delta_N-\hbar\omega)\nonumber,\\
  \gamma^{-}_N (\omega)=\sum_{r=R,L}\gamma^{-}_{r}(\Delta_{N-1}+\hbar\omega),
\end{eqnarray}
all the diagrams in Fig.~\ref{fig:irreducible} can be calculated in
the same way, resulting in

\begin{equation}
  \begin{array}{lll}
    a)\Rightarrow \gamma^{+}_N(\omega), & d)\Rightarrow \gamma^{-}_N(\omega), & g)\Rightarrow-\gamma^{-}_N(\omega)\\
    b)\Rightarrow \gamma^{+}_N(-\omega), & e)\Rightarrow \gamma^{-}_N(-\omega), & h)\Rightarrow-\gamma^{+}_N(\omega)\\
    c)\Rightarrow-\gamma^{-}_N(-\omega),& f)\Rightarrow
    -\gamma^{+}_N(-\omega) & i)\Rightarrow\frac{i}{\omega}.\\
  \end{array}
\end{equation}

In the same way, the vertex corrections in Fig.~\ref{fig:all_vertex}
can be calculated yielding
\begin{widetext}
\begin{equation}
  \label{eq:vertex}
  \begin{array}{lll}
    a) \frac{i}{\omega}[\gamma^{+}_{N}(0)-\gamma^{+}_{N}(\omega)], &
    d) \frac{i}{\omega}[\gamma^{-}_{N}(0)-\gamma^{-}_{N}(\omega)], &
    g) -\frac{i}{\omega}[\gamma^{-}_{N}(0)-\gamma^{-}_{N}(\omega)] \\
    b) \frac{i}{\omega}[\gamma^{+}_{N}(0)-\gamma^{+}_{N}(-\omega)], &
    e) \frac{i}{\omega}[\gamma^{-}_{N}(0)-\gamma^{-}_{N}(-\omega)], &
    h) -\frac{i}{\omega}[\gamma^{+}_{N}(0)-\gamma^{+}_{N}(\omega)]\\
    c) -\frac{i}{\omega}[\gamma^{-}_{N}(0)-\gamma^{-}_{N}(-\omega)], &
    f) -\frac{i}{\omega}[\gamma^{+}_{N}(0)-\gamma^{+}_{N}(-\omega)], &
    i) 1,\\
  \end{array}
\end{equation}
\end{widetext}
where the expression in Eq.~(\ref{eq:vertex}$i)$ corresponds to the zeroth order correction of the vertex.
To lowest order, the total spectral density can be written
\begin{eqnarray}
  \label{eq:sv_firstorder}
  S_V(\omega)&=&2\frac{e^2}{C^2} \mathrm{Re} \left\{ \sum_{N,N^\prime,N^{\prime\prime}}  P^{st}_{N}
     \hat{V}_{N,N^\prime}(\omega)  \hat{\Pi}_{N^\prime,N^{\prime\prime}}(\omega)
    N^{\prime\prime}\right\}\nonumber \\
  &=& 2\frac{e^2}{C^2} \mathrm{Re} \left\{ \overrightarrow{N}^{T}
    \hat{\Pi}(\omega) \hat{V}(\omega) \overrightarrow{P}^{st} \right\}
\end{eqnarray}
where $\overrightarrow{N}$ and $\overrightarrow{P}^{st}$ are column
vectors ($\overrightarrow{N}^{T}$ is the transpose of $\overrightarrow{N}$) containing the number of
extra electrons on the island and the steady-state probabilities
respectively. Note thus that the element $P_{0}^{st}$ refers to the steady-state
probability of the SET to be in the state with $0$ extra charges,
unlike for the qubit, where $P_{0}^{st,qb}$ refers to the steady-state
probability of being in the lower energy eigenstate. $\hat{V}(\omega)$ is a matrix whose elements $\hat{V}_{N^\prime,N}(\omega)$ are the sum of all
vertex corrections in Fig.~\ref{fig:all_vertex} which take the system from the
state $N$ to the state $N^\prime$ and $\hat{\Pi}(\omega)$ is a matrix
whose elements $\hat{\Pi}_{N^{\prime\prime},N^{\prime}}(\omega)$ are the sum of all diagrams
in Fig.~\ref{fig:irreducible} that take the system from the
state $N^\prime$ to the state $N^{\prime\prime}$.
Incidentally, Eq.~(\ref{eq:sv_firstorder}) is valid for arbitrary
order, as long as the vertex correction and the propagator are dressed
to the appropriate order.

\begin{figure}[htbp]
  \begin{center}
    \includegraphics[width=6cm]{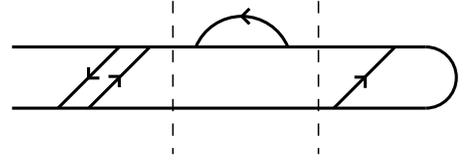}
    \caption{This diagram contains three irreducible diagrams as
      in between it is possible to draw a vertical auxiliary line
      (dashed lines) separating them.}
    \label{fig:irreducibles}
  \end{center}
\end{figure}

In order to facilitate the evaluation of higher order diagrams we
define an irreducible diagram as a diagram where it is impossible to
draw an auxiliary vertical line at any time, without crossing an electron-hole line.
An example can be seen in Fig.~\ref{fig:irreducibles}.

The diagrams in Fig.~\ref{fig:irreducible} are all the first order
irreducible diagrams, except Fig.~\ref{fig:irreducible}i), which is just the free propagator
\mbox{$\hat{\Pi}^{(0)}_{N,N^\prime}(\omega)=\frac{i}{\omega}\delta_{N,N^\prime}$},
where $\delta_{N,N^\prime}$ is a Kronecker delta.

\begin{figure}
  \begin{center}
    \includegraphics[width=8cm]{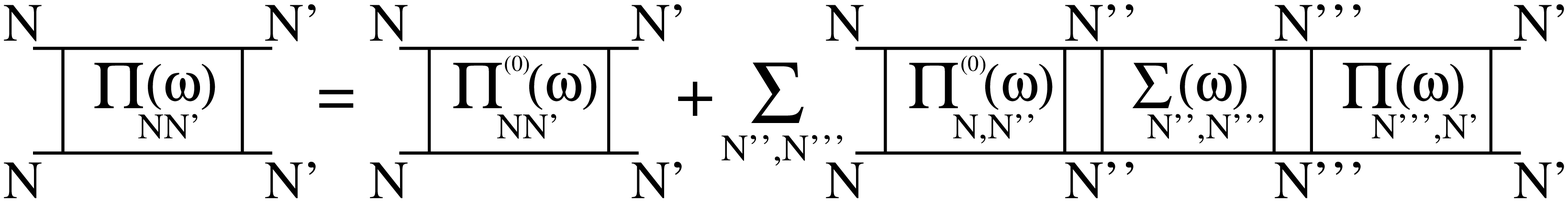}
    \caption{A graphical representation of the Dyson equation in
      Eq.~(\ref{eq:dyson}). The terms $\Sigma_{NN^{\prime}}$ in are
      irreducible diagrams that take the system from the state $N$ to
      the state $N^{\prime}$.}
    \label{fig:dyson}
  \end{center}
\end{figure}

Using irreducible diagrams allows us to write down a matrix Dyson equation in frequency space
for the frequency dependent propagator $\hat{\Pi}(\omega)$ between different charge states  (Fig.~\ref{fig:dyson})

\begin{equation}
  \label{eq:4}
  \hat{\Pi}(\omega) = \hat{\Pi}^{(0)}(\omega) + \hat{\Pi}^{(0)}(\omega)\hat{\Sigma}(\omega)\hat{\Pi}(\omega).
\end{equation}
Solving for $\hat{\Pi}(\omega)$ and inserting the explicit form of
$\hat{\Pi}^{(0)}(\omega)$ we get
\begin{equation}
  \label{eq:dyson}
  \hat{\Pi}(\omega)=\frac{i}{\omega}\left(\mathbf{1}-\frac{i\hat{\Sigma}(\omega)}{\omega}\right)^{-1}.
\end{equation}
Note that the explicit time ordering in every diagram, means that in
frequency space any specific diagram can be written as a product of irreducible
diagrams and free propagators.

The matrix element $\hat{\Sigma}_{N^\prime,N}(\omega)$ of the
irreducible propagator (self-energy) $\hat{\Sigma}(\omega)$ is the sum of the
irreducible diagrams that take the system from the state $N$ to the state $N^\prime$
and $\mathbf{1}=\delta_{N,N^\prime}$ 
is the unit matrix with the same dimension as $\hat{\Sigma}(\omega)$.
Note that the frequency dependence comes directly from the Laplace transform
over the time $\tau$, which introduces an auxiliary line with energy $\hbar\omega$.
All diagrams located between the times $t=\tau$ and $t=0$ therefore depend on $\omega$.

\subsection{The self-energy $\hat{\Sigma}(\omega)$}
The Dyson equation allows us to appropriately sum up the diagrams to a
certain order by calculating the self energy to that order and then
inserting it into Eq.~(\ref{eq:dyson}).

As the reservoirs are assumed to be in local
equilibrium, we chose to include only diagrams containing at most
one electron-hole excitation at any given time. This approximation
corresponds to keeping the irreducible diagrams where any vertical
line cuts at the most one internal line. This is equivalent to the sequential tunneling
approximation leading to the Master equation of orthodox
SET-theory.

The diagrams entering the self-energy to this order are all drawn in Fig.~\ref{fig:irreducible}a)-h).

\subsection{Vertex Corrections}
To calculate the noise spectral density we also need to sum all vertex
corrections to the same order.

As the sequential tunneling approximation only includes terms with at
 most one tunneling event at a time, all the vertex corrections
that enter are those drawn in Fig.~\ref{fig:all_vertex}.

\subsection{Main Result}
\begin{figure*}
  \begin{center}
    \includegraphics[width=\textwidth]{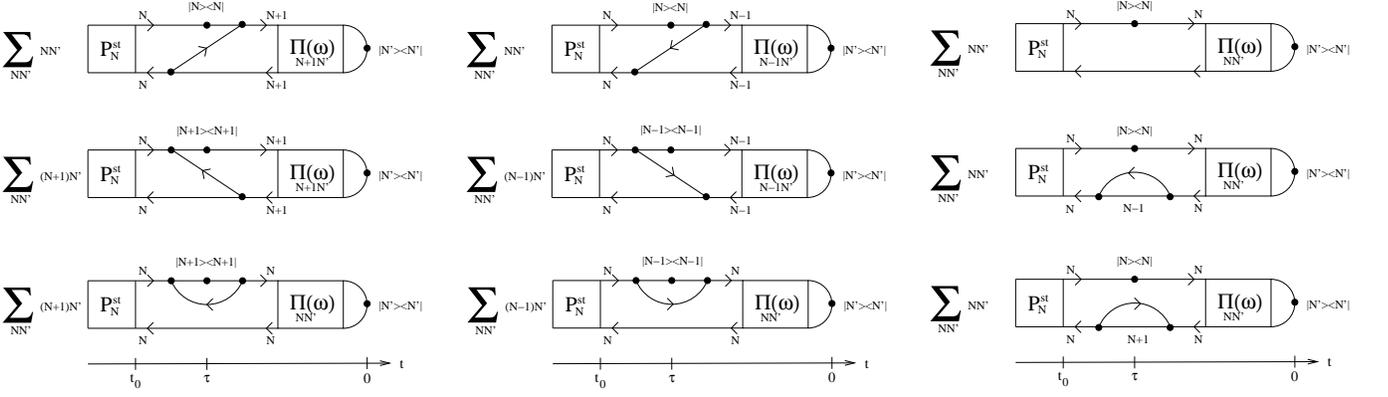}
    \caption{The sum of all diagrams within the sequential tunneling
      approximation. The propagators $\Pi_{NN^{\prime}}$, which is the
      sum over all the diagrams that take the system from the state
      $N$ to the state $N^{\prime}$, are calculated using a Dyson
      equation (drawn graphically in Fig.~\ref{fig:dyson}).}
    \label{fig:Sdiag}
  \end{center}
\end{figure*}

All diagrams which enter Eq.~(\ref{eq:newS}) within our approximations are drawn in Fig.~\ref{fig:Sdiag}.
Adding them together gives
\begin{widetext}
  \begin{eqnarray}
    \label{eq:fullS}
    S_{V}(\omega)&=&\frac{2e^2}{C^2} \sum_{N} P_{N}^{st} \mathrm{Re}
    \left\{ N \Lambda_{N}-\frac{i}{\omega}
    \left[(N+1)\gamma_{N}^{+}(\omega)+N\gamma_{N}^{+}(-\omega))\right](\Lambda_{N+1}-\Lambda_{N})-\nonumber\right.\\
    &&\left.-\frac{i}{\omega}
    \left[(N-1)\gamma^{-}_N(\omega)+N\gamma_{N}^{-}(-\omega)\right](\Lambda_{N-1}-\Lambda_{N})\right\},
  \end{eqnarray}
\end{widetext}

where $\Lambda_{N}=\sum_{N^{\prime}} N^{\prime}
\Pi_{N^{\prime},N}(\omega)$ and $P_{N}^{st}$ is the steady-state
probability of there being $N$ extra electrons on the island.
In Eq.~(\ref{eq:fullS}) we have used that the steady-state
probabilities fulfill the relation
$P_{N+1}^{st}=\frac{\gamma_{N}^{+}(0)}{\gamma_{N+1}^{-}(0)}P^{st}_{N}$
(see for instance \cite{ProbandRand}).

The first term in Eq.~(\ref{eq:fullS}) corresponds to the result in
Eq.~(27) of Ref.~\cite{Korotkov_spectral} but with frequency dependent
tunneling rates, while the other terms originate from the
vertex corrections. 
The asymmetry in the
noise with respect to positive and negative frequencies arises solely
from these vertex corrections.
In the limit $\omega \rightarrow 0$ Eq.~(\ref{eq:fullS}) coincides
with the zero frequency limit given in Ref.~\cite{Korotkov_spectral}.
Note that the frequency independent terms in the
vertex corrections cancel.

Using the relation $\mathrm{Re} \{\hat{\Pi}(\omega)\} = -
\frac{1}{\omega}\hat{\Sigma}(\omega) \mathrm{Im}
\{\hat{\Pi}(\omega)\}$ (this is the real part of
Eq.~(\ref{eq:4}), using that the free propagator and the first order self energies are purely imaginary) Eq.~(\ref{eq:fullS}) can be rewritten in matrix
form, which gives the main result in this paper,
\begin{equation}
  \label{eq:s_full_matrix}
  S_{V}(\omega)=\frac{2e^2}{C^2}
  \overrightarrow{N}^{T}\left[\mathbf{1}+\left(\frac{\hat{\Sigma}(\omega)}{\omega}\right)^2\right]^{-1}\frac{\hat{\gamma}(\omega)}{\omega^2} \overrightarrow{P}^{st},
\end{equation}
where $\hat{\gamma}(\omega)$ is a tridiagonal matrix whose elements
are
\mbox{$\hat{\gamma}_{N,N}(\omega)=\gamma_{N}^{-}(\omega)-\gamma_{N}^{+}(\omega)$},
\mbox{$\hat{\gamma}_{N+1,N}(\omega)=\gamma_{N}^{+}(\omega)$} and \mbox{$\hat{\gamma}_{N-1,N}(\omega)=-\gamma_{N}^{-}(\omega)$}.

Note that contrary to $\hat{\Sigma}(\omega)$, $\hat{\gamma}(\omega)$
is not symmetric in frequency and for negative frequencies larger than
the maximally available energy from the SET $\hat{\gamma}(\omega)$ is
analytically zero while for positive frequencies,
$\hat{\gamma}(\omega)$ does not tend to zero but to a finite value.

\section{Back-action during measurement}
\label{sec:back-action}
When measuring with the SET, the bias voltage is typically large enough
to allow for a DC-current through the SET but not much larger, which
implies that only the charge states $0$ and $1$ have a non-zero
steady-state probability. In this case they are given by
$P_{0}^{st}=\frac{\gamma_{1}^{-}(0)}{\gamma_{1}^{-}(0)+\gamma_{0}^{+}(0)}$
and $P_{1}^{st}=\frac{\gamma_{0}^{+}(0)}{\gamma_{1}^{-}(0)+\gamma_{0}^{+}(0)}$.
We assume $\Delta_{0}^{L}<0$,
$\Delta_{0}^{R}>0$ and $|\Delta_{0}^{L}| > |\Delta_{0}^{R}|$ so that
electrons typically tunnel from the left to the right, and there
is a finite DC-current through the SET. We will use these assumptions
about the bias throughout the remaining part of the paper, except
in the section about the off-state noise of the normal state SET.

\subsection{Low-frequency regime}
In the low-frequency regime, defined as the regime where $\gamma_0^-(\omega)$
and $\gamma_1^+(\omega)$ are exponentially small,
the charge states $0$ and $1$ are the only states energetically accessible,
also taking into account the externally available energy
$\hbar\omega$. 
In this case the matrix inversion in Eq.~(\ref{eq:s_full_matrix}) is
easy to calculate analytically and the noise spectral density is given by
\begin{equation}
  S_V (\omega)=
  \frac{2e^2}{C^2}
  \frac{P_{0}^{st}\gamma_{0}^{+}(\omega)+P_{1}^{st}\gamma_{1}^{-}(\omega)}{\omega^2+[\gamma_{0}^{+}(\omega)+\gamma_{0}^{+}(-\omega)+\gamma_{1}^{-}(\omega)+\gamma_{1}^{-}(-\omega)]^2}.
  \label{eq:low_freq}
\end{equation}
This expression has a very simple form: The sum of the steady state
probabilities weighted by the inelastic tunneling rates for
transitions away from the state, normalized by a denominator
containing the finite lifetimes of the states. 
For zero frequency this corresponds to classical telegraph-noise.
Note that Eq.~(\ref{eq:low_freq}) is valid both in the
normal and superconducting states, the difference only entering
in the expressions for the rates $\gamma_{0,1}^{\pm}(\omega)$.

\subsection{High-frequency regime}
In the high frequency limit the spectral noise density of the SET
should be independent of the bias and be dominated by the Nyquist
noise, which in this regime ($\hbar\omega \gg k_B T$) is~\cite{CallenWellton} 
\begin{equation}
  \label{eq:nyquist}
  S^{Nyq}_{V}(\omega)=2 \hbar\omega \mathrm{Re}\left\{Z(\omega)\right\}=
\frac{2\hbar\omega R_{||}}{1+\left(\omega R_{||}C\right)^2},
\end{equation}
where $Z(\omega)$ is the impedance of the SET island to ground and
$R_{||}=\left(1/R_T^L+1/R_T^R\right)^{-1}$. 
In this limit $\hat{\Sigma}(\omega) \ll \omega$ and the matrix
inversion in Eq.~(\ref{eq:s_full_matrix}) can be Taylor expanded and
approximated by the first term
$\frac{i}{\omega}\left[\mathbf{1}+\left(\frac{\hat{\Sigma}(\omega)}{\omega}\right)^2\right]^{-1}\approx
\frac{i}{\omega}\mathbf{1}$. Still assuming the voltage bias to be small enough to keep only the steady-state
probabilities $P_{0}^{st}$ and $P_{1}^{st}$ non-zero,
 Eq.~(\ref{eq:s_full_matrix}) gives
\begin{equation}
\label{eq:highfreqnoise}
  S_V(\omega)=\frac{2e^2}{C^2} \frac{P_{0}^{st}[\gamma_{0}^{+}(\omega)+\gamma_{0}^{-}(\omega)]+P_{1}^{st}[\gamma_{1}^{-}(\omega)+\gamma_{1}^{+}(\omega)]}{\omega^2}.
\end{equation}

In the high frequency limit, $\hbar\omega \gg \{E_C, eV\}$, all rates are similar
and they are proportional both to the normal state tunnel conductance and the frequency
\begin{equation}
\label{eq:highfreqrates}
\gamma_{0}^{\pm}(\omega)\approx\gamma_{1}^{\pm}(\omega)=
\frac{\hbar\omega}{2e^2}\left[\frac{1}{R_T^L}+\frac{1}{R_T^R}\right]+O(1),
\end{equation}
where $O(1)$ indicates a bias-dependent constant. This is valid both in the normal
and superconducting states. It is clear that inserting Eq.~(\ref{eq:highfreqrates})
into Eq.~(\ref{eq:highfreqnoise}) gives Eq.~(\ref{eq:nyquist}).
It might be interesting to note that it is enough to include four
charge states to recover the full Nyquist noise.
If only two charge states were included, an extra charge on the island
would prevent further electrons to tunnel until the extra electron has
left the island, and the correlation effectively would reduce the noise to that of a single junction.
For similar reasons, for intermediate frequencies the noise should be reduced, compared to the
Nyquist noise.

\subsection{Normal State SET}
 For an SET operated in the normal state, the density of states can be
assumed to be energy independent when calculating the tunneling rates in
Eq.~(\ref{eq:gammarp}). Using
\mbox{$\rho_{I,n}(E)=\rho_{r,n}(E)=\rho_N$}, 
the tunneling rates $\gamma^{\pm}(\omega)$ can be written
\begin{equation}
  \label{eq:normalrateplus}
  \gamma^{+}_{N}(\omega)=\frac{\pi}{\hbar} \sum_r \alpha_{0}^{r} \Gamma^{+}(\Delta_{N}^{r}-\hbar\omega)
\end{equation}
\begin{equation}
  \label{eq:normalrateminus}
  \gamma^{-}_{N}(\omega)=\frac{\pi}{\hbar} \sum_r \alpha_{0}^{r} \Gamma^{-}(\Delta_{N-1}^{r}+\hbar\omega)
\end{equation}
where $\alpha_{0}^{r}=\sum_{n}|T^{rn}|^2\rho_{N}^2=
\frac{R_k}{4\pi^2 R_{T}^{r}}$ is the dimensionless conductivity ($R_k$
is the quantum resistance and $R_{T}^{r}$ is the tunneling resistance
of junction $r$), 
\mbox{$\Gamma^{-}(E)=E/(1-\exp(-\beta E))$}, \mbox{$\Gamma^{+}(E)=E \exp(-\beta
E)/(1-\exp(-\beta E))$} and $\beta=1/k_B T$. Note that $\Delta_{N}^{r}$
includes both charging and biasing energies.

At zero temperature, the $\Gamma^{\pm}$ become step functions
multiplied by a linear term, 
\mbox{$\Gamma^{-}(E)= |E|\theta(E)$}, \mbox{$\Gamma^{+}(E)= |E|\theta(-E)$}. In this
limit the rates are easy to analyze.

\begin{figure}
  \begin{center}
    \includegraphics[width=7cm]{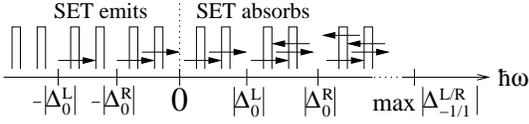}
    \caption{A schematic picture of the different processes in the SET
      for different frequencies. Note that negative frequencies
      correspond to processes where the SET emits energy, while
      positive frequencies correspond to processes where the SET
      absorbs energy.}
    \label{fig:freq_proc}
  \end{center}
\end{figure}

\begin{figure}
  \begin{center}
    \includegraphics[width=8cm]{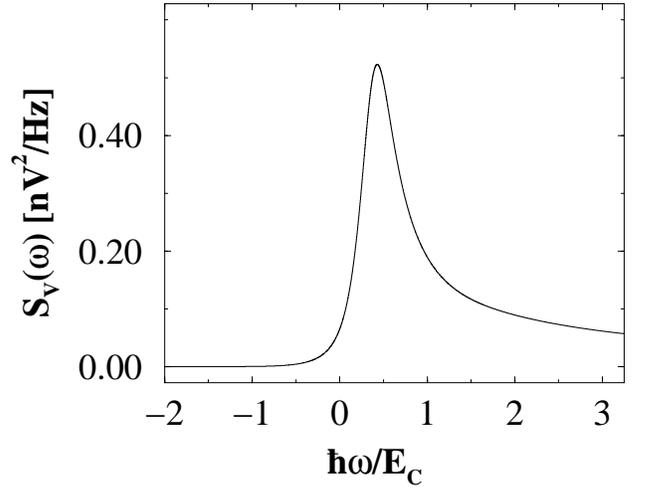}
    \caption{Spectral noise density for an SET run in normal mode. The
      calculation was done for zero temperature and with symmetric
      tunnel junctions $R_R=R_L=21.5$ k$\Omega$. The DC-current through
      the SET was 1.5nA, $n_x=0.25$, \mbox{$E_C=2.5$ K}.}
    \label{fig:S_low}
  \end{center}
\end{figure}

For frequencies of small magnitudes
$|\hbar\omega|<|\Delta_{1}^{r}|,|\Delta_{-1}^{r}|$ only the charge states $\ket{0}$ and
$\ket{1}$ are energetically allowed and we can use
Eq.~(\ref{eq:low_freq}) to calculate the noise spectral density.
Even though the expression in Eq.~(\ref{eq:low_freq}) looks
very much like the classical expression with frequency dependent rates,
this frequency dependence of the rates changes the behaviour quite drastically. The
spectral noise density is no longer symmetric with respect to
$\omega$, and there is a finite maximum energy available for emission
from the SET, which can be seen in Fig.~\ref{fig:S_low} as
$S_V=0$ for large negative frequencies $\hbar\omega <
-|\Delta_{0}^{L}|$. This means that if the energy splitting of the qubit is larger than the energy gained
by putting an extra electron on the island ($\Delta_{0}^{L}$), there
isn't enough energy available from the SET to excite the qubit, and the
SET behaves as a passive load, only able to absorb energy.

The preference of the SET to absorb energy
rather than emit is also clear as $S_V(\omega) > S_V(-\omega)$ for
any $\omega >0$. This means that any two-level system with finite
energy splitting driven to steady-state solely by the SET will not
have an equal steady-state probability of both states. 

\begin{figure}
  \begin{center}

    \includegraphics[width=8cm]{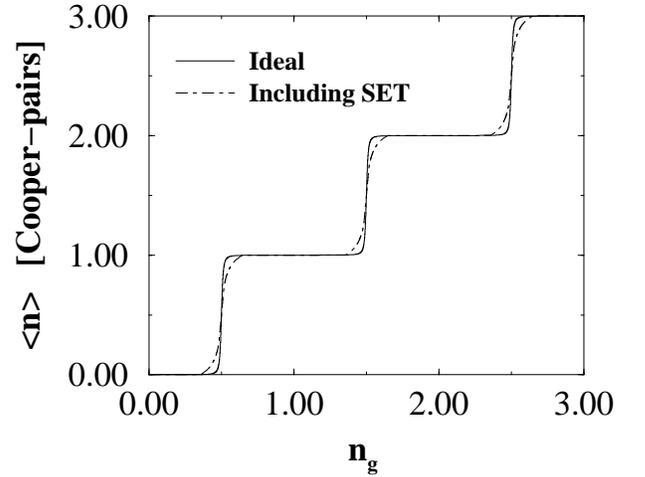}
    \caption{The Coulomb staircase of an SCB driven to steady state by
      the SET run in normal state. The parameters we have used for the SCB are
      $E_{qb}=2.5$ K, $E_J=0.1$ K, $\Delta=2.5$ K and for the SET we have
      used $E_C=2.5$ K, \mbox{$R_R=R_L=21.5$ k$\Omega$} and
      $n_x=0.25$.}
    \label{fig:normalstairs}
  \end{center}
\end{figure}

\subsubsection{Low-frequency regime}
For low frequencies $|\hbar\omega| < |\Delta_{0}^{R}|$ when no backward
tunneling processes are allowed, the noise spectral density can be
written
\begin{equation}
  \label{eq:Slow}
  S_V(\omega)=\frac{e^2}{C^2}\frac{2I/e+2\pi\omega\left[P_{0}^{st}\alpha_{0}^{L}+P_{1}^{st}\alpha_{0}^{R}\right]}{\omega^2+4\left[\gamma_{0}^{+}(0)+\gamma_{1}^{-}(0)\right]^2},
\end{equation}
where the first term in the numerator \mbox{$I=2 e
\gamma^{+}_{0}(0)\gamma^{-}_{1}(0)/[\gamma^{+}_{0}(0)+\gamma^{-}_{1}(0)]$}
is the DC-current though the SET. We see that the difference compared with classical
telegraph-noise is the term linear in $\omega$ in the nominator. This is a
quantum mechanical correction originating from the vertex corrections. 
In this regime the frequency dependent part of the
tunneling rates in the denominator cancel.

In the symmetrised noise $S_V^{sym}(\omega)=S_V(\omega)+S_V(-\omega)$, the linear term 
in the numerator cancels out. Thus in this region, for quantities that are proportional
to the symmetrized noise, such as the mixing-time (see Eq.~(\ref{eq:mixtime})), the classical telegraph-noise
give the same result. But for other quantities, such as the steady-state
probabilities of a qubit driven by the SET (see Eq.~(\ref{eq:steadyqubitup}) and
Eq.~(\ref{eq:steadyqubitdown})), the difference is 
evident even for small frequencies.

\subsubsection{Coulomb staircase}
Using the SET to measure the average charge of the Cooper-pair box qubit
it is reasonable to assume that the back-action from the SET is the dominant
noise source.
At the degeneracy point of the qubit, the energy splitting between its two 
eigenstates is $E_J$. If $E_J < \Delta_{0}^{R}$ we can use Eq.~(\ref{eq:Slow}) to
calculate the Coulomb staircase (Eq.~\ref{eq:coulomb}) close to the degeneracy as
\begin{equation}
  \label{eq:1}
  \left< Q \right> =e\left[1+ \frac{8 E_c \pi \alpha_0}{\hbar I/e} \delta n_g\right]=
e\left[1+ \frac{4 E_c}{e R_T I} \delta n_g\right],
\end{equation}
where $\delta n_g$ is the deviation from the degeneracy point
($n_g=1/2$) and we have assumed symmetric junctions
($\alpha_{0}^{L}=\alpha_{0}^{R}=\alpha_{0}$ or $R_T^L=R_T^R=R_T$)
and a symmetric voltage bias in the SET. 
Thus, close to the degeneracy, we will always get a linear charge
increase for suitable choice of SET bias.
In this regime the derivative
is thus determined by the current through the SET rather than the
Josephson energy in the qubit.

Away from the degeneracy point, when the energy
splitting of the qubit is increased, the low-frequency requirement
for Eq.~(\ref{eq:Slow}) may not be fulfilled.
In order to calculate the influence from the noise on the Coulomb
staircase for arbitrary qubit gate voltage we have
to include the full expression from Eq.~(\ref{eq:fullS}). The
result for a typical setup is plotted in
Fig.~\ref{fig:normalstairs}, demonstrating that the back-action noise from the SET
introduces additional smearing of the Coulomb staircase. 

This can be compared to the results by Nazarov \cite{NazarovStairs},
where the influence of the back-action of an SET in the normal state
is calculated on a small metallic island in the normal state.

\subsubsection{Mixing time}
\begin{figure}[h]
  \begin{center}
    \includegraphics[width=8cm]{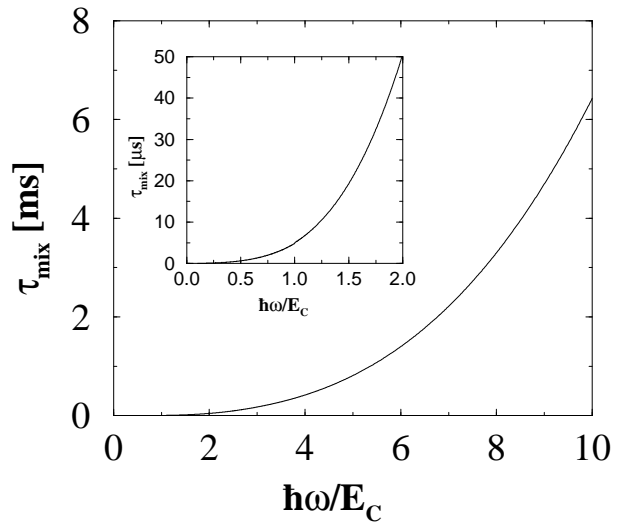}
    \caption{The mixing time of an SCB caused by the SET. The inset
      shows an expanded view around $\hbar \omega \approx E_{qb}$. Thus, for an
      energy splitting of the qubit of approximately $E_{qb}$, the mixing
      time is around 10 $\mu$s. The parameters used were $E_C=2.5$ K,
      $E_{qb}=0.8$ K, $E_J=0.15$ K, $R_R=R_L=21.5$ k$\Omega$,
      $\kappa=0.01$, $I_{DC}=9.6$ nA.}
    \label{fig:mixingnormal}
  \end{center}
\end{figure}

Using the tunneling rates in
Eq.~(\ref{eq:normalrateplus},\ref{eq:normalrateminus})                                                        
and inserting them into Eq.~(\ref{eq:fullS}) we can calculate the mixing time $\tau_{mix}$
due to the voltage fluctuation on the SET-island as a function of the
energy splitting. Using a state-of-the-art RF-SET\cite{AassimeAPL}
coupled to a qubit with realistic parameters (see caption), as shown
in Fig.~\ref{fig:mixingnormal}, this would give a mixing
time of approximately $10$ $\mu$s. This should be compared with the
measurement time $t_{ms}$ needed to resolve the two charge states in the same
setup which is about $0.4 \mu$s. The resulting signal-to-noise
ratio (SNR) is $SNR=\sqrt{\tau_{mix}/t_{ms}}\approx 5$, which indicates
that single-shot read-out is possible.

\subsubsection{Off-state noise - Qubit reset}
One property of the SET used as a charge qubit read-out device is that
it may be switched off by lowering the driving bias so that sequential
tunneling is no longer possible, i.e. both $0<\Delta_0^L$ and $0<\Delta_0^R$.
In this regime the voltage noise is determined by
co-tunneling processes\cite{SomeOldCoTunnelReference}.
Since co-tunneling is a second order process in the tunneling conductance
the voltage noise in the off-state
\cite{AlecMaasen,Averin} is several orders of magnitude smaller than the on-state noise.

\begin{figure}[h]
  \begin{center}
    \includegraphics[width=8cm]{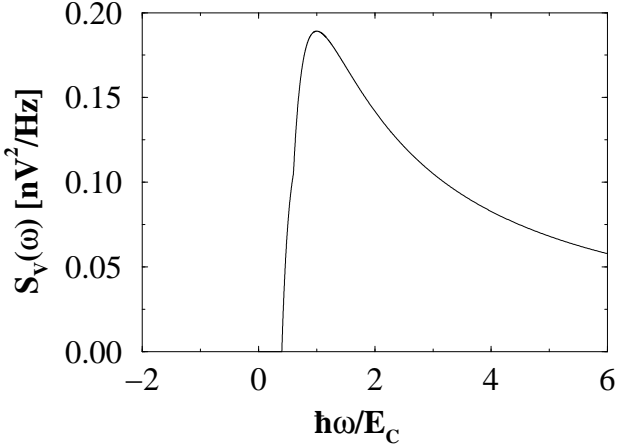}
    \caption{Noise spectral density of an SET in the off-state. Note
      that only contributions from positive frequencies remain, as no
      energy can be emitted from the SET within the sequential
      tunneling approximation. The noiseless region 
      is given by $\Delta_{0}^{l}-\hbar \omega >0$.}
    \label{fig:normaloff}
  \end{center}
\end{figure}

Taking energy exchange with the qubit into account there may be
a first order tunneling event in the SET, even though the driving bias
is too small for sequential tunneling. The energy taken from relaxing
the qubit may stimulate a photon-assisted first-order tunnel event in
the SET. At zero temperature the condition for such an event is
simply $\Delta E > \min\{\Delta_{0}^{L},\Delta_{0}^{R}\}$.
The voltage noise spectral density of the SET in the off-state
is shown in Fig.~\ref{fig:normaloff}. The curve has been calculated
using Eq.~(\ref{eq:fullS}), with $P_0^{st}=1$.

This implies that in order to benefit from the low voltage fluctuations
in the off-state the SET should be switched off by switching both
the driving bias to zero and using the SET gate voltage to put it far
into the Coulomb co-tunneling regime, i.e $n_x\approx 0$.

The nonlinearity of the voltage noise spectral density may also be
used for fast relaxation of the qubit, i.e. as a qubit reset button.
If the gate voltage of the SET is such that
$\Delta E \geq |\Delta_{0}^{L}| \approx \Delta_{0}^{R}$, and the driving
bias is zero, the qubit relaxation rate is first-order in the tunnel
conductance, while the excitation rate is given by co-tunneling.
The normal state SET may thus be used for qubit reset, or in other
words as a switchable dissipative environment to the qubit.

\subsection{Superconducting SET}

Compared with a normal state SET (NSET) the superconducting SET (SSET) shows two main
differences. The density of states in the reservoirs is changed by the
superconducting energy gap $\Delta$, and in addition to quasiparticle
tunneling also Cooper pair tunneling may occur.
We will consider an SSET biased so that sequential quasiparticle
tunneling is allowed, and in this regime Cooper pair tunneling
may be neglected. Thus the same model as before can be used,
only taking into account the changed quasiparticle density of states.
As we are interested in an SET made out of aluminum, we use the
BCS density of states
\begin{equation}
  \rho(E)=\rho_N \frac{|E|}{E^2-\Delta^2}\theta(|E|-\Delta),
\end{equation}
where $\rho_N$ is the density of states of the normal state.
Inserting these into the expression for the tunneling rates in
Eq.~(\ref{eq:gammarp}) we get for zero temperature (see e.g. Ref.~\cite{Tinkham})
\begin{widetext}
\begin{eqnarray*}
  \label{eq:SratePT0}
  \gamma^{\pm}_{r}(w)&=&\frac{\pi}{\hbar}\frac{\alpha^{r}_{0} \theta(\mp\hbar \omega
  \pm eV_r-2 \Delta)}{2 \Delta\mp\hbar\omega\pm eV_r} {\Bigg [} (\hbar\omega-eV_r)^2 \mathbf{K}
  (\frac{\hbar\omega - eV_r\pm 2\Delta}{\hbar\omega-eV_r\mp
    2\Delta})-\\
  & & -(2\Delta\mp \hbar\omega\pm eV_r)^2\left\{\mathbf{K}(\frac{\hbar\omega -eV_r\pm
  2\Delta}{\hbar\omega-eV_r\mp 2\Delta})-\mathbf{E}(\frac{\hbar\omega
  -eV_r\pm 2\Delta}{\hbar\omega-eV_r\mp 2\Delta})\right\} {\Bigg ]},
\end{eqnarray*}
\end{widetext}
where $\mathbf{K}(x)$ and $\mathbf{E}(x)$ are elliptic integrals of
the first and second kind. These rates behave just as the IV curve for
an SIS-junction.
The singularities in the superconducting density of states introduce discontinuities into the
tunneling rates. 
These discontinuities will also introduce discontinuities in the
noise spectral density.

\begin{figure}
  \begin{center}
    \includegraphics[width=8cm]{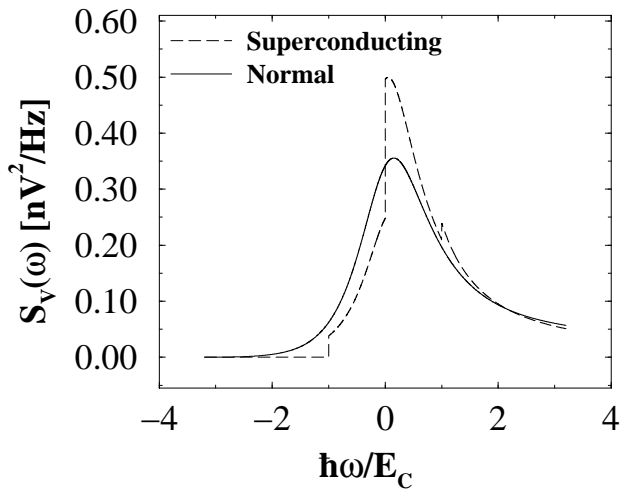}
    \caption{Comparing the Spectral noise density for a
      superconducting and a normal conducting SET. The parameters used
      were $R_{L}=R_{R}=21.5$ k$\Omega$, $E_C=2.5$ K, $n_x=0.25$ and
      $I_{DC}=9.6$ nA.\label{fig:compare}}
  \end{center}
\end{figure}

\subsubsection{Comparison between an SSET and an NSET}
Comparing the noise spectral density of an NSET and an SSET (see Fig.~\ref{fig:compare})
is not completely straightforward as the SSET
requires considerably higher voltage bias in order to get sequential
quasiparticle tunneling
through the SET, i.e. $|eV_L-eV_R|>4\Delta+E_C(1-2n_x)$.
Therefore when comparing these two
in the on-state (i.e. while measuring),
we use the same tunnel conductance and gate voltage, and then choose
a voltage bias that gives the same DC-current through the two SETs. This is motivated
by the fact that the zero frequency noise is determined by the DC-current through
the SET, this biasing therefore yields the same zero frequency telegraph noise for both
the SSET and the NSET.

Apart from the discontinuities in the spectral density of the SSET, the finite frequency noise
differs in another important aspect. Although the two SETs carry
the same DC current, the processes producing that current are
qualitatively different. In the superconducting SET biased just
above the threshold the energy gain in each single tunnel event
is quite small, determined by approximately $\max\{|eV_L|,|eV_R|\}-2\Delta$.
The relatively large current is an effect of the divergent density
of state peaks in the reservoirs. In the normal state SET carrying
the same current the maximum energy that may be extracted from a
single tunneling event is instead quite large, proportional to $\max\{|eV_L|,|eV_R|\}$.

Comparing the voltage noise spectral density for negative frequencies,
capable of exciting the measured system, we find that the SSET
noise is zero for $\hbar\omega < -(\max\{|eV_L|,|eV_R|\}-2\Delta)$,
while the NSET spectrum extends down to
$\hbar\omega \approx -\max\{|eV_L|,|eV_R|\}\approx -2\Delta$.

\begin{figure}[htbp]
  \begin{center}
    \includegraphics[width=8cm]{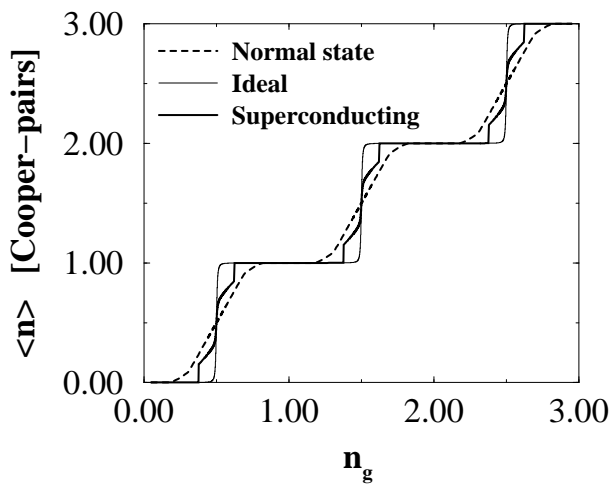}

    \caption{Comparison of an ideal Coulomb staircase and a staircase
      where the qubit is driven to steady state by either an SET in
      the normal or in the superconducting state. We have use the same
      parameters as in Fig.~\ref{fig:normalstairs}.}
    \label{fig:metastairs}
  \end{center}
\end{figure}

Measuring the Coulomb staircase with an NSET and an SSET biased to
the same DC-current will thus give different results.
The Coulomb staircase is sharper for the SSET because
the lower amount of energy extractable from the SET reduces the
excitation rate for the two-level system, and the discontinuities in the
noise spectral density of the SSET are also clearly visible, as
seen in Fig.~\ref{fig:metastairs}.
Even though this is a completely different bias regime, similar
structure appears in Ref.\cite{GirvinJQPnoise}.

Note that the staircases in Fig.~\ref{fig:metastairs} has been
calculated for zero temperature and for a fixed voltage bias across
the SET, and that the DC-current is different in
Fig.~\ref{fig:metastairs} and Fig.~\ref{fig:normalstairs}.

When calculating the total mixing time, the sum of relaxation and
absorption rates enters, and the difference between an SSET and an
NSET diminishes. The lower tendency for the superconducting SET to emit is
compensated for by an increased tendency to absorb energy.
\begin{figure}[htbp]
  \begin{center}
    \includegraphics[width=8cm]{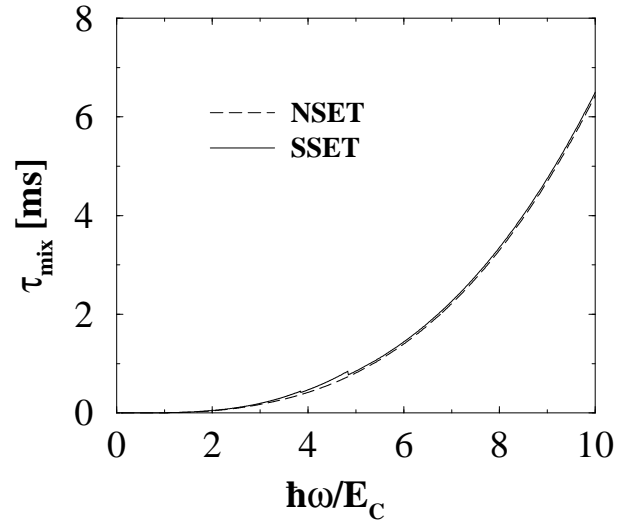}
    \caption{The mixing time due to the noise from the SET run either
      in the normal state or in the superconducting state, using the
      same DC-current through the SET (approximately 10 nA).}
    \label{fig:sumnoise}
  \end{center}
\end{figure}

Since the mixing time due to an SSET dependends on the
sum of the contributions from absorptive and emissive processes, it is
thus not very different from an NSET carrying the same DC-current.
An example can be seen in Fig.~\ref{fig:sumnoise}.

\section{Conclusions}
We have calculated the full frequency spectral density of voltage fluctuations in a
Single Electron Transistor (SET), used as an electrometer biased above the
Coulomb threshold so that the current through the SET is carried by sequential
tunneling
 events. We take the energy exchange between the SET and the measured system
into account using a real-time diagrammatic Keldysh technique. We find simple
analytical expressions for the noise in the low- and high-frequency
regimes and in between we calculate the noise numerically. The complexity
of the numerical calculation is limited to the inversion of a $N X N$ matrix where
$N$ is the number of charge states involved, typically $N \leq 5$.

Previous expressions for the voltage fluctuations, where the energy exchange
is not taken into account, are by definition symmetric with respect to positive and
negative frequencies. We show that there is an asymmetry, technically arising from
the first order vertex corrections of the external vertices, so that the noise capable of exciting
the measured system is always less than the noise that will relax the
measured system, at any given frequency. The importance of this difference
is shown by calculating the Coulomb staircase of a Cooper pair box, as
measured by the SET. Interestingly the difference has a tendency to cancel
in the expression for the symmetric noise, i.e. the sum of the positive and negative
frequency noise. This implies that the classical calculation is a reasonably good
approximation for that quantity.

The divergence in the superconducting density of states results in discontinuities
in the voltage noise spectral density of the superconducting SET (SSET). Compared to a
normal state SET carrying the same DC current the SSET also has considerably less ability
to excite the measured system.

\section{Acknowledgments}
We gratefully acknowledge rewarding discussions with Per Delsing, David Gunnarsson, Kevin Bladh,
Vitaly Shumeiko, Alexander Zazunov, Alexander Shnirman and Yuriy
Makhlin. 
This work was partially funded by the Swedish grant
agency NFR/VR and by the SQUBIT project of the IST-FET
programme of the EC.

\end{document}